\documentclass[twocolumn,showpacs]{revtex4}
\usepackage{graphicx}
\begin{document}
\title{Broad histogram relation for the bond number 
and its applications}

\author{Chiaki Yamaguchi}
\affiliation{
Department of Physics, Tokyo Metropolitan University,
Hachioji, Tokyo 192-0397, Japan
}
\author{Naoki Kawashima}
\affiliation{
Department of Physics, Tokyo Metropolitan University,
Hachioji, Tokyo 192-0397, Japan
}

\author{Yutaka Okabe}
\affiliation{
Department of Physics, Tokyo Metropolitan University,
Hachioji, Tokyo 192-0397, Japan
}
\date{\today}

\begin{abstract}
We discuss Monte Carlo methods based on the cluster (graph) 
representation for spin models.  We derive a rigorous broad histogram 
relation (BHR) for the bond number; a counterpart for the energy 
was derived by Oliveira previously.  A Monte Carlo dynamics 
based on the number of potential moves for the bond number is proposed.  
We show the efficiency of the BHR for the bond number 
in calculating the density of states and other physical quantities. 
\end{abstract}


\maketitle

\section{Introduction}

The development of new algorithms for the Monte Carlo simulation 
is important to overcome the problem of slow dynamics.  
We may classify such attempts into two categories.  
The first category is the extended ensemble method; 
one uses an ensemble different from the ordinary 
canonical ensemble with a fixed temperature. 
The multicanonical method \cite{Berg92,Lee93}, 
the simulated tempering \cite{Marinari92}, the exchange Monte Carlo 
method \cite{Hukushima96}, the broad histogram method \cite{Oliv96}, 
the flat histogram method \cite{Wang98,Wang00}, and the Wang-Landau 
algorithm \cite{WL01} are examples of the first category.  
The second category includes the cluster algorithm;  
one flips a large number of spins in a correlated cluster 
at a time instead of a single-spin flip, which helps 
the relaxation time decrease drastically. 
Examples of the second category are 
the Swendsen-Wang (SW) algorithm \cite{SW87} and the Wolff 
algorithm \cite{Wolff89}.  
Recently Tomita and Okabe \cite{PCC} proposed 
an effective cluster algorithm, which is called 
the probability-changing cluster algorithm, 
of tuning the critical point automatically.  

The combination of approaches of two categories is a challenging 
problem to explore an efficient algorithm. 
Janke and Kappler \cite{Janke} proposed a trial to combine 
the multicanonical method and the cluster algorithm; 
their method is called the multibondic ensemble method. 
Quite recently, Yamaguchi and Kawashima \cite{Yama02} have 
improved the multibondic ensemble method; they have also shown 
that the combination of the Wang-Landau algorithm and 
the improved multibondic ensemble method yields much better statistics 
compared to the original multibondic ensemble method 
by Janke and Kappler \cite{Janke}.  

One calculates the energy density of states (DOS) $g(E)$ 
in the multicanonical method \cite{Berg92,Lee93} and 
the Wang-Landau method \cite{WL01}; 
the energy histogram $H(E)$ is checked 
during the Monte Carlo process. 
In contrast, the DOS for bond number $n_b$, 
$\Omega (n_b )$, is calculated in the multibondic ensemble 
method \cite{Janke} or the improved multibondic ensemble method 
by Yamaguchi and Kawashima \cite{Yama02}; the histogram 
for bond number, $H(n_b)$, is checked in the Monte Carlo 
process.  

In proposing the broad histogram method, 
Oliveira {\it et al.}~\cite{Oliv96} paid attention to 
the number of potential moves, or the number of 
the possible energy change, $N(S, E \to E')$, 
for a given state $S$.  The total number of moves is
$$
 \sum_{\Delta E} N(S, E \to E + \Delta E) = N
$$
for a single-spin flip process, where $N$ is the number of spins. 
The energy DOS is related to the number of potential moves as
\begin{equation}
 g(E) \, \left< N (S, E \to E') \right>_E  
   = g(E') \, \left< N (S', E' \to E) \right>_{E'}, 
\label{BHR}
\end{equation}
where $\left< \cdots \right>_E$ denotes the microcanonical 
average with fixed $E$.  This relation is shown to be valid 
on general grounds \cite{Oliv98}, and hereafter we call 
Eq.~(\ref{BHR}) as the broad histogram relation (BHR) 
for the energy. 
One may use the number of potential moves $N(S, E \to E')$ 
for the probability of updating states.  While 
the original dynamics \cite{Oliv96} was criticized to be 
not entirely correct \cite{Wang98,Berg98}, a refined 
dynamics is employed in the flat histogram method \cite{Wang00}. 
Alternatively, one may employ other dynamics which has no relation 
to $N(S, E \to E')$, but Eq.~(\ref{BHR}) 
is used when calculating the energy DOS \cite{Oliv00,Lima00}. 
It was stressed \cite{Oliv00,Lima00} that $N(S, E \to E')$ 
is a macroscopic quantity, which is the advantage of using 
the number of potential moves.  We do not have to care about the relative 
number of visits for different energy level $E$.  It is contrary 
to the case of the multicanonical method \cite{Berg92,Lee93} 
or the Wang-Landau method \cite{WL01}. 
The only crucial point is the uniformity of visits within 
the same energy level \cite{Oliv00}.

It is quite interesting to ask whether there is a relation 
similar to the BHR, Eq.~(\ref{BHR}), for the bond number.  
In this paper, using the cluster (graph) representation, 
we derive the BHR for the bond number.  
We propose a dynamics based on the number of potential moves 
for the bond number.  Using the DOS 
for the bond number thus obtained, we calculate the specific heat 
for model spin systems.  We also employ other dynamics, that is, 
the multibondic ensemble method \cite{Janke} and 
its improvement \cite{Yama02}, and calculate the 
bond-number DOS based on the BHR for the bond number. 
Comparing the efficiency of several methods, we show that 
the calculation of the bond-number DOS through 
the BHR gives much better statistics 
compared to the direct calculation of the DOS. 

The rest of the paper is organized as follows.  In Sec. II, 
we briefly review the cluster (graph) representation  
for the $Q$-state Potts model.  In Sec. III, we derive 
the BHR for the bond number.  A dynamics 
based on the number of potential moves for the bond number 
is discussed in Sec. IV.  
In Sec. V, calculating the accuracy of the specific heat 
for the two-dimensional (2D) Ising model, we compare 
the efficiency of several methods.  
The summary and discussions are given in Sec. VI. 

\section{Cluster formalism}

We briefly review the cluster (graph) formalism for the $Q$-state 
Potts model.  We are concerned with the Hamiltonian
$$  
{\cal H} = - J \sum_{\left< i,j \right>} \delta_{\sigma_i, \sigma_j},
  \quad \sigma_i = \{ 1, \cdots , Q\}
$$
where $J$ is the exchange coupling constant and 
the summation is taken over the nearest-neighbor pairs $\left< i,j \right>$. 
From now on, we represent the energy in units of $J$, 
and the Boltzmann constant is set to be one. 

The partition function for a given temperature $T$ is expressed as
$$ 
 Z(T) \equiv \sum_{S} W_0(S) = \sum_{E} g(E) \, W_0(E(S),T)
$$
with the Boltzmann weight of state $S$ having the energy $E$,
$$
 W_0(S) = W_0(E(S),T) = e^{-E(S)/T},
$$
and the energy DOS,
$$
 g(E) \equiv \sum_{\{ S|E(S)=E \}} 1.
$$

With the framework of the dual algorithm \cite{KD,KG}, 
the partition function is also expressed in the double summation 
over state $S$ and graph $G$ as 
$$
Z (T) = \sum_{S, G} V_0 (G) \, \Delta (S, G),
$$
where $\Delta (S, G)$ is a function that takes the value one 
when $S$ is compatible to $G$ and takes the value zero otherwise. 
A graph consists of a set of bonds.  
The weight for graph $G$, $V_0(G)$, is defined as
$$
  V_0(G) = V_0(n_b(G),T) = (e^{1/T} - 1)^{n_b(G)} 
$$
for the $Q$-state Potts model, where $n_b(G)$ is 
the number of ``active'' bonds in $G$.
This is nothing but the Fortuin-Kasteleyn
representation \cite{FK1} for the $Q$-state Potts model. 
We say a pair $(i, j)$ is satisfied if $\sigma_i = \sigma_j$, 
and unsatisfied otherwise.
Satisfied pairs become active with a probability $p = 1-e^{-1/T}$ 
for given $T$.

By taking the summation over $S$ and $G$ with fixing the number 
of bonds $n_b$, 
the expression for the partition function becomes 
$$
Z (T) = \sum_{n_b = 0}^{N_B} \Omega (n_b) \, V_0 (n_b, T), 
$$
where $N_B$ is the total number of nearest-neighbor pairs 
in the whole system.
Here, $\Omega (n_b )$ is the DOS for the bond number defined as
the number of consistent combinations of
graphs and states such that the graph consists of $n_b$ bonds;
$$
\Omega (n_b) \equiv \sum_{\{G|n_b (G) = n_b\}} \sum_S \, \Delta (S, G).
$$

Then, the canonical average of a quantity $A$ is calculated by
\begin{equation}
\left< A \right>_T = 
\frac{\sum_{n_b} \left< A \right>_{n_b} \Omega (n_b ) 
 V_0 (n_b, T)}{Z (T)}, 
\label{eq:canonical}
\end{equation}
where $\left< A \right>_{n_b}$ is 
the microcanonical average with the fixed bond number $n_b$
for the quantity $A$ defined as 
\begin{equation}
 \left< A \right>_{n_b} \equiv 
 \frac{ \sum_{ \{G|n_b (G) = n_b \} } \sum_S A(S, G)
 \Delta (S, G)}{\Omega (n_b)}.
\label{eq:micronb}
\end{equation}
Thus, if we obtain $\Omega(n_b)$ and $\left< \cdots \right>_{n_b}$ 
during the simulation process, we can calculate the canonical 
average of any quantity. 

We should note that for the calculation of the energy $E$, 
it is convenient to use the relation 
\begin{equation}
\left< E \right>_T = T^2 \frac{d}{d T} \log Z (T)
 = - \frac{e^{1/T}}{e^{1/T} - 1} \, \left<n_b \right>_T.
\label{eq:energy}
\end{equation}
Similarly, the specific heat per one site $C$ is given by
\begin{eqnarray}
 C N T^2  &=& - \frac{e^{1/T}}{(e^{1/T} - 1)^2} \left< n_b \right>_T
 \nonumber \\
 & & + \biggl( \frac{e^{1/T}}{e^{1/T} - 1} \biggr)^2 
 (  \left< n_b^2 \right>_T - \left< n_b \right>_T^2 ). 
\label{eq:specific}
\end{eqnarray}
The above equations (\ref{eq:energy}) and (\ref{eq:specific}) 
were derived by Janke and Kappler \cite{Janke}.

\section{BHR for the bond number}

The relation between the energy DOS and the number of potential moves 
for energy, the BHR for the energy, was rigorously derived by 
Oliveira \cite{Oliv98}.  
Here we follow a method similar to that used 
by Oliveira to derive the BHR for the bond number. 
Instead of using the relation between states,
we consider the relation between graphs.  

The number of potential moves from the graph 
with the bond number $n_b$ to the graph with $n_b+1$, 
$N(S, G, n_b \to n_b+1)$, for fixed $S$ is equal to 
that of the number of potential moves 
from the graph with $n_b+1$ to that with $n_b$, 
$N(S, G', n_b + 1 \to n_b)$.  That is, the following relation 
is satisfied:
\begin{eqnarray}
&~& 
\sum_{\{G|n_b(G) = n_b \}} N(S,G,n_b \to n_b+1) =
 \nonumber \\
 &~& \quad \quad
\sum_{\{G'|n_b(G') = n_b+1 \}} N(S,G',n_b+1 \to n_b). \nonumber \\
\label{eq:bro1}
\end{eqnarray}
Taking a summation over states $S$ and
using the definition of the microcanonical average 
with the fixed bond number $n_b$, Eq.~(\ref{eq:micronb}), 
we rewrite Eq.~(\ref{eq:bro1}) as
\begin{eqnarray}
&~& \Omega (n_b) \left< N(G, n_b \to n_b + 1) \right>_{n_b} \nonumber \\
 &~& \quad = \Omega (n_b + 1) \, \left< N(G', n_b + 1 \to n_b) \right>_{n_b + 1}.
\label{eq:broad}
\end{eqnarray}
This is the BHR for the bond number.
It should be noted that $N(G, n_b \to n_b + 1)$ is a possible 
number of bonds to add, and related to the number of satisfied pairs 
for the given state $S$,
$$
 n_p(S) = \sum_{\left< i,j \right>} \delta_{\sigma_i (S), \sigma_j (S)},
$$
by
$$
 N(G, n_b \to n_b+1) = n_p(S) - n_b. 
$$
With use of the microcanonical average with fixed bond number 
for $n_p$, we have the relation 
\begin{equation}
 \left< N(G, n_b \to n_b + 1) \right>_{n_b}
 = \left< n_p \right>_{n_b} - n_b.
\label{eq:transition1}
\end{equation}
On the other hand, the possible number of bonds to delete, 
$N(G', n_b + 1 \to n_b)$, is simply given by $n_b+1$, that is,
\begin{equation}
 \left< N(G', n_b + 1 \to n_b) \right>_{n_b + 1} = n_b + 1.
\label{eq:transition2}
\end{equation}
From the BHR for the bond number, Eq.~(\ref{eq:broad}), we have
\begin{equation}
 \frac{\Omega(n_b)}{\Omega(0)} = 
  \prod_{l=0}^{n_b-1} \frac{\Omega(l+1)}{\Omega(l)} =
  \prod_{l=0}^{n_b-1}
  \frac{\left< N(G, l \to l+1) \right>_{n_b=l}}
       {\left< N(G, l+1 \to l) \right>_{n_b=l+1}}
\label{eq:broad2}
\end{equation}
Then, substituting Eqs.~(\ref{eq:transition1}) 
and (\ref{eq:transition2}) into Eq.~(\ref{eq:broad2}), we obtain 
the bond-number DOS, $\Omega (n_b)$, as 
\begin{equation}
 \ln \Omega (n_b) = \ln \Omega (0) + \sum_{l=0}^{n_b - 1}
 \ln \biggl(
 \frac{\left< n_p \right>_{n_b=l} - l}{l + 1}
 \biggr).
\label{eq:Ocal}
\end{equation}
When calculating the bond-number DOS from the BHR for 
the bond number, we only need the information on 
$\left< n_p \right>_{n_b}$, the microcanonical average 
with fixed $n_b$ of the number of satisfied pairs $n_p$. 
It is much simpler than the case of the BHR formulation 
for the energy DOS.  

Moreover, in the computation of $n_p$, we can use 
an improved estimator. If a pair of sites $(i,j)$ belong to 
the different cluster, this pair is satisfied with a probability 
of $1/Q$.  If a pair of sites belong to the same cluster, 
this pair is always satisfied.
Then, we can employ an improved estimator $\tilde{n}_p$ as 
\begin{equation}
\tilde{n}_p(G)  = \biggr( 1 - \frac{1}{Q} \biggl)
 \sum_{\left< i,j \right>} \delta_{c_i(G), c_j(G)} + \frac{N_B}{Q},
\label{eq:np2}
\end{equation}
where $c_i(G)$ represent a cluster that a site $i$ belongs to. 
Only the information on graph is needed.  By definition, 
$\left< \tilde{n}_p \right>_{n_b} = \left< n_p \right>_{n_b}$.
We employ the improved estimator in the whole calculation below.
Inserting Eq.~(\ref{eq:np2}) into Eq.~(\ref{eq:Ocal}), we have 
\begin{equation}
 \frac{\Omega (n_b)}{Q^N} = \frac{1}{n_b !}
 \prod_{l = 0}^{n_b - 1} \biggl[
 (1 - \frac{1}{Q}) \left<  \delta_{c_i (G), c_j (G)} \right>_{n_b=l} 
  + \frac{N_B}{Q} - l
\biggr].
\label{eq:Ocal2}
\end{equation}
Here we have used the relation 
$$
 \Omega(0) = Z(T \to \infty) = Q^N.
$$
It is interesting to check Eq.~(\ref{eq:Ocal2}) for a special case. 
The $Q \to 1$ limit of the $Q$-state Potts model 
is the bond percolation problem. 
If we substitute $Q = 1$ into Eq.~(\ref{eq:Ocal2}), 
we obtain 
$$
\Omega (n_b) = \left( \begin{array}{c} N_B \\ n_b \end{array} \right), 
$$
which is the expected relation for the bond percolation problem. 

\section{Flat histogram method for the bond number}

Let us consider the update process for the Monte Carlo 
simulation.  
In the multibondic ensemble method, a graph is updated 
by adding or deleting a bond for a satisfied pair 
of sites \cite{Janke}. 
The histogram $H(n_b)$ becomes flat 
if we use the following rule. 
If there is a bond already on the chosen pair, 
we delete it with a probability
\begin{equation}
 P(n_b \to n_b - 1) = \frac{\Omega(n_b)}{\Omega(n_{b-1})+\Omega(n_b)}, 
\label{eq:prob1}
\end{equation}
On the other hand, if there is no bond and if the pair is satisfied, 
we add a bond with a probability
\begin{equation}
 P(n_b \to n_b + 1) = \frac{\Omega(n_b)}{\Omega(n_{b+1})+\Omega(n_b)}.
\label{eq:prob2}
\end{equation}
Since the exact form of the bond-number DOS $\Omega(n_b)$ 
is not known {\it a priori}, we renew $\Omega(n_b)$ iteratively 
in the Monte Carlo process by several ways \cite{Janke,Yama02}.  

We may use the number of potential move for the bond number, 
$\left< N (G, \cdots ) \right>_{n_b}$, for the probability of update. 
Inserting Eqs.~(\ref{eq:broad}), (\ref{eq:transition1}), and 
(\ref{eq:transition2}) into Eqs.~(\ref{eq:prob1}) and (\ref{eq:prob2}), 
we get the probability to delete a bond, 
\begin{equation}
 P(n_b \to n_b - 1) = \frac{\left< n_p \right>_{n_b - 1} + 1 - n_b}
{\left< n_p \right>_{n_b - 1} + 1}, 
\label{eq:prob3}
\end{equation}
and the probability to add a bond, 
\begin{equation}
P(n_b \to n_b + 1) = \frac{n_b + 1}{\left< n_p \right>_{n_b} + 1}, 
\label{eq:prob4}
\end{equation}
respectively.

The actual Monte Carlo procedure is as follows.  
We start from some state (spin configuration) $S$, and 
an arbitrary graph $G$ consistent with it.  We add or delete 
a bond of satisfied pairs with the probability 
(\ref{eq:prob3}) or (\ref{eq:prob4}). 
After making such a process as many as the number of total pairs, $N_B$, 
we flip every cluster 
with the probability 1/2.  And we repeat the process. 
Since we do not know the exact form of $\left< n_p \right>_{n_b}$, 
we use the accumulated average for $\left< n_p \right>_{n_b}$. 
The dynamics proposed here can be regarded as 
the flat histogram method for the bond number, which we call 
the cluster-flip flat histogram method.  The conventional 
flat histogram method for the energy \cite{Wang00} will be referred to 
as the single-spin-flip flat histogram method hereafter. 
As $\left< n_p \right>_{n_b}$ converges to the exact value, 
the histogram $H(n_b)$ becomes flat.  We calculate 
the bond-number DOS by using Eq.~(\ref{eq:Ocal2}), 
and then calculate various quantities by Eq.~(\ref{eq:canonical}), 
or Eqs.(\ref{eq:energy}) and (\ref{eq:specific}).

Here, we have described the procedure for the multiple cluster 
update of the Swendsen-Wang type \cite{SW87}, 
but we can also employ the single cluster update 
of the Wolff type \cite{Wolff89}. 

\section{Results}

\begin{figure}
\includegraphics[width=0.95\linewidth]{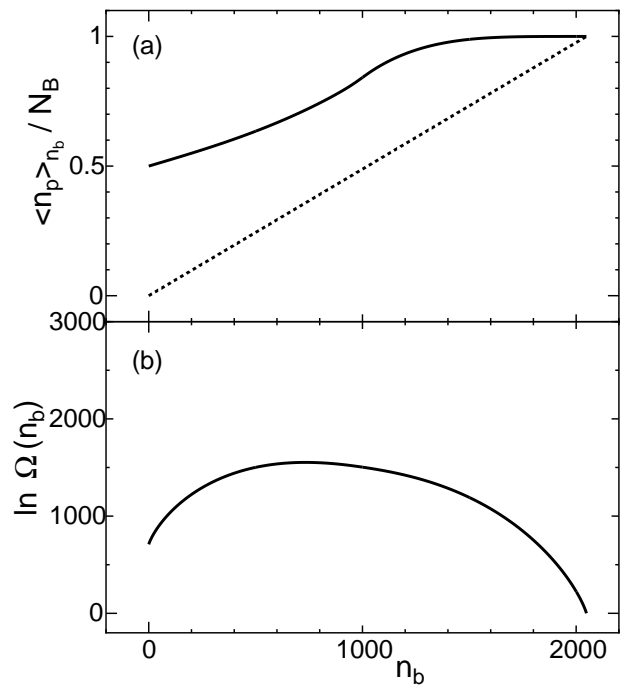}
\caption{
(a) $\left< n_p \right>_{n_b}/N_B$ and (b) $\ln \Omega(n_b)$ 
of the $32 \times 32$ Ising model 
obtained by the cluster-flip flat histogram method.  
The dotted line in (a) denotes $n_b/N_B$.}
\label{fig:1}
\end{figure}

\begin{figure}
\includegraphics[width=0.95\linewidth]{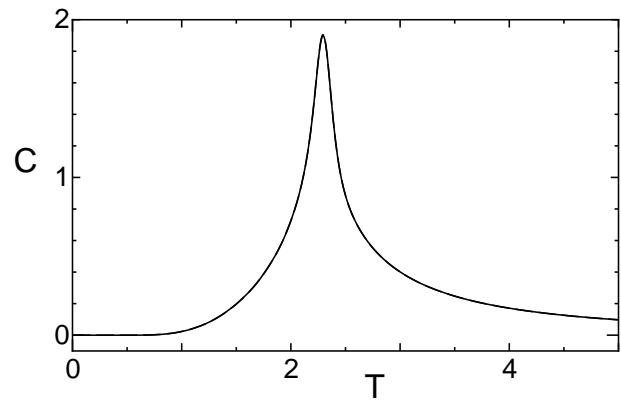}
\caption{Specific heat per a site of the 2D Ising model for $L$ = 32 
obtained by the cluster-flip flat histogram method.}
\label{fig:2}
\end{figure}

First, we simulate the $L \times L$ Ising model on the square lattice 
with the periodic boundary conditions by using 
the cluster-flip flat histogram method. 
We show $\left< n_p \right>_{n_b}/N_B$ as a function of $n_b$ for $L=32$ 
by the solid line in Fig.~\ref{fig:1}(a); we give $n_b/N_B$ by the dotted line. 
The number of Monte Carlo sweeps (MCS) is $5 \times 10^7$. 
The difference between the solid and dotted lines represents the number of 
potential moves $\left< N(n_b \to n_b+1) \right>/N_B$, whereas the difference 
between the dotted line and the horizontal axis represents 
$\left< N(n_b \to n_b-1) \right>/N_B$.  We should note that 
$\left< n_p \right>_{n_b=0}/N_B = 1/2$, which is expected from Eq.~(\ref{eq:np2}). 
The logarithm of the bond-number DOS, $\ln \Omega(n_b)$, obtained 
by $\left< n_p \right>_{n_b}$ is shown in Fig.~\ref{fig:1}(b) as a function of $n_b$.  
The temperature dependence of the specific heat 
calculated using Eq.~(\ref{eq:specific}) is shown in Fig.~\ref{fig:2}; 
the deviation from the exact result obtained by Beale \cite{Beale} 
is not visible in this scale. 

\begin{figure}
\includegraphics[width=0.95\linewidth]{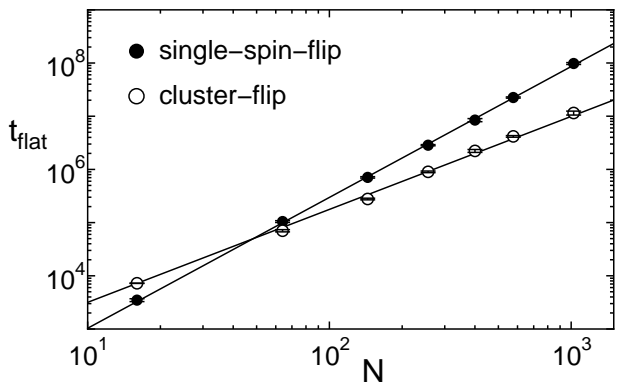}
\caption{Size dependence of the flatness time for the 2D Ising model. 
The linear system sizes $L$ are 4, 8, 12, 16, 20, 24, and 32; $N=L^2$.
The cluster-flip flat histogram method and the single-spin-flip 
flat histogram method are compared.}
\label{fig:3}
\end{figure}

Let us compare the performance of the cluster-flip 
flat histogram method proposed in this paper with that of 
the single-spin-flip flat histogram method \cite{Wang00}.  
To do this, we check the number of MCS to satisfy 
the flatness condition for the histogram $H(n_b)$ or $H(E)$; 
we state that the flatness condition is fulfilled if the histogram $H(n_b)$ 
or $H(E)$ for all possible $n_b$ or $E$ is equal to or larger than 80\% of 
the average histogram $\overline{H}$.  In Fig.~\ref{fig:3}, 
we show the size dependence of the number of MCS to satisfy 
the flatness condition, which we call the flatness time 
$t_{\rm flat}$ hereafter, 
for both the cluster-flip flat histogram method and 
the single-spin-flip flat histogram method in logarithmic scale.  
The linear system sizes $L$ are 4, 8, 12, 16, 20, 24, and 32.  
The average is taken over many samples.  The number of samples 
ranges from 20 for the largest system to 1000 for the smallest. 
We see from Fig.~\ref{fig:3} that for the single-spin-flip 
flat histogram method the flatness time increases more rapidly 
as the system size increases.  The least-squares fitting of the data gives
$$
 \ln t_{\rm flat} \sim 4.04(2) + 1.75(1) \times \ln N
$$
for the cluster flat histogram method, and 
$$
 \ln t_{\rm flat} \sim 1.28(7) + 2.46(1) \times \ln N
$$
for the single-spin-flip flat histogram method. 

\begin{figure}
\includegraphics[width=0.95\linewidth]{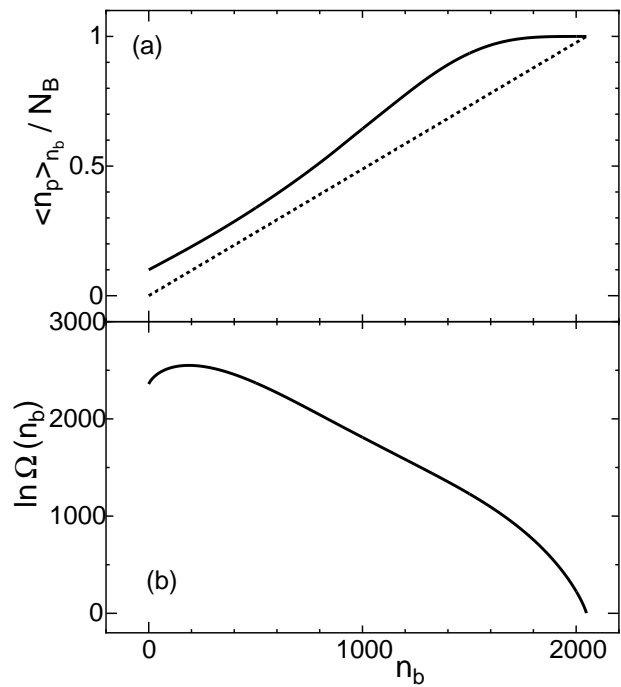}
\caption{
(a) $\left< n_p \right>_{n_b}/N_B$ and (b) $\ln \Omega(n_b)$ 
of the $32 \times 32$ 10-state Potts model 
obtained by the cluster-flip flat histogram method. 
The dotted line in (a) denotes $n_b/N_B$.}
\label{fig:4}
\end{figure}

\begin{figure}
\includegraphics[width=0.95\linewidth]{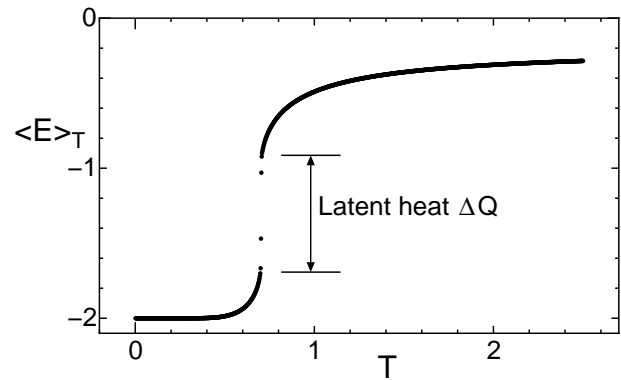}
\caption{Energy of the 2D 10-state Potts model for $L$ = 32 
obtained by the cluster-flip flat histogram method.}
\label{fig:5}
\end{figure}

\begin{figure}
\includegraphics[width=0.95\linewidth]{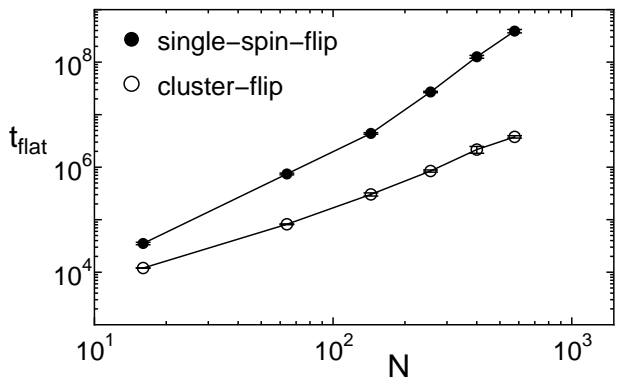}
\caption{Size dependence of the flatness time for the 2D 10-state Potts model. 
The linear system sizes $L$ are 4, 8, 12, 16, 20, and 24; $N=L^2$.
The cluster-flip flat histogram method and the single-spin-flip 
flat histogram method are compared.}
\label{fig:6}
\end{figure}

As another example, we simulate the 2D 10-state Potts model 
on the square lattice.  A strong first-order phase transition occurs 
in this model.  
We show $\left< n_p \right>_{n_b}/N_B$ for the $32 \times 32$ lattice 
by the solid line in Fig.~\ref{fig:4}(a); we give $n_b$ by the dotted line. 
The number of MCS is $5 \times 10^7$. 
The number of potential moves $\left< N(n_b \to n_b+1) \right>/N_B$ and 
$\left< N(n_b \to n_b-1) \right>/N_B$ are given in the same manner as 
the case of the Ising model.  It is to be noted that 
$\left< n_p \right>_{n_b=0}/N_B = 1/10$ for the 10-state Potts model. 
The logarithm of the bond-number DOS, $\ln \Omega(n_b)$, obtained by 
$\left< n_p \right>_{n_b}$ is shown in Fig.~\ref{fig:4}(b). 
The temperature dependence of the energy obtained by 
Eq.~(\ref{eq:energy}) is given in Fig.~\ref{fig:5}. 
The latent heat $\Delta Q$ is shown in the figure. 
The comparison of the flatness time for the 2D 10-state 
Potts model is shown in Fig.~\ref{fig:6}. 
The linear system sizes $L$ are 4, 8, 12, 16, 20, and 24.  
The number of samples to take the average 
ranges from 5 for the largest system to 1000 for the smallest. 
The flatness time of the single-spin-flip flat histogram method 
increases more rapidly with size than that of 
the cluster-flip flat histogram method, although 
it is not clear whether the size dependence is linear or not 
in logarithmic scale.  It again shows the superiority of 
the cluster-flip flat histogram method over the single-spin-flip 
flat histogram method. 

\begin{figure}
\includegraphics[width=0.9\linewidth]{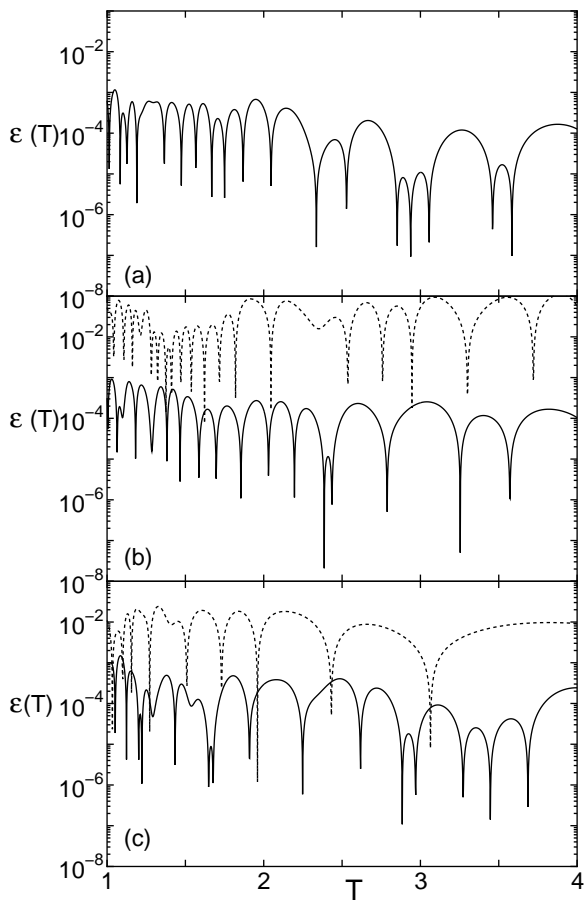}
\caption{Relative errors of the specific heat for the $32 \times 32$ 
Ising model; 
(a) the cluster-flip flat histogram method, 
(b) the multibondic ensemble method, and 
(c) the improved multibondic ensemble method.  
The number of MCS is $20000 \times N_B$. 
The solid line denotes the data obtained by the calculation 
using the number of potential moves, and dotted line denotes those obtained 
by the direct calculation with $H(n_b)$. }
\label{fig:7}
\end{figure}

In the calculations presented above, we have used the number of 
potential moves both for the dynamics and the estimator 
of $\Omega(n_b)$ or $\left< n_p \right>_{n_b}$. 
However, our procedure to calculate the bond-number DOS $\Omega(n_b)$ 
using the number of potential moves, or more explicitly, 
using $\left< n_p \right>_{n_b}$, 
Eq.~(\ref{eq:Ocal}) or (\ref{eq:Ocal2}), is independent of the dynamics. 
We may use the multibondic ensemble method \cite{Janke} or 
its improvement \cite{Yama02}, and monitor $\left< n_p \right>_{n_b}$ 
to compute $\Omega(n_b)$, 
although $\Omega(n_b)$ is directly used for the probability to update 
and renewed with the help of the histogram $H(n_b)$, 
such as $\Omega^{\rm old}(n_b)H(n_b) \to \Omega^{\rm new}(n_b)$. 
We compare the accuracy of the calculation for several dynamics and 
the procedure to calculate $\Omega(n_b)$. 
For that purpose, we study the errors of the specific heat 
for the 2D Ising model.  The energy DOS is exactly calculated 
by Beale \cite{Beale}.  As already shown in Fig.~\ref{fig:2}, 
the errors of our calculation are very small; 
we treat the relative error, which is defined as 
$$
\epsilon(T) \equiv \biggl|
\frac{C_{{\rm simulation}}(T) - C_{{\rm exact}}(T)}{C_{{\rm exact}}(T)}
\biggr|,
$$
for the specific heat $C$. 
The relative errors $\epsilon (T)$ of the $32 \times 32$ Ising model 
in the case of the cluster-flip flat histogram method 
are shown in Fig.~\ref{fig:7}(a).  
The number of MCS is $20000 \times N_B$. 
The average value of $\epsilon(T)$ in the range of $1.0 \le T \le 4.0$,  
which will be denoted by $\overline{\epsilon (T)}$, is 
as small as 0.0002. 

In the case of the multibondic ensemble method, we can calculate 
$\Omega(n_b)$ either through the number of potential moves or by the direct 
calculation with the help of the histogram $H(n_b)$.  
The errors $\epsilon(T)$ of the $32 \times 32$ Ising model 
in the case of the multibondic ensemble method are plotted 
in Fig.~\ref{fig:7}(b).  The number of MCS is $20000 \times N_B$; 
we renew $\Omega(n_b)$ for the probability of graph update 
by every $100 \times N_B$ MCS.  The solid line denotes the data 
for the calculation using the number of potential moves, and 
dotted lines denotes those for the direct calculation using $H(n_b)$. 
We see that 
the calculation of $\Omega(n_b)$ through the number of potential moves gives 
much smaller errors.  The average value $\overline{\epsilon (T)}$ is 
0.0002 for the calculation using the number of potential moves, whereas 
that for the direct calculation with $H(n_b)$ is 0.043.  
We also show the results of the improved multibondic method 
in Fig.~\ref{fig:7}(c).  The conditions are the same as those 
for the multibondic method.  The average value $\overline{\epsilon (T)}$ 
for the calculation using the number of potential moves is 0.0002, whereas 
that for the direct calculation with $H(n_b)$ is 0.0087.  
The calculation of $\Omega(n_b)$ through the number of potential moves again 
gives much smaller errors compared to the direct calculation 
with $H(n_b)$. 
It is interesting to notice that $\overline{\epsilon (T)}$ take 
almost the same value for several methods if we follow the procedure 
to calculate $\Omega(n_b)$ through $\left< n_p \right>_{n_b}$. 
The data of $\overline{\epsilon (T)}$ for several methods 
are tabulated in Table \ref{table1} for convenience.
\begin{table}
\caption{Average relative error of the specific heat $\overline{\epsilon (T)}$ 
for the 2D $32 \times 32$ Ising model.  We compare the data 
for several Monte Carlo methods 
and the procedure to calculate $\Omega(n_b)$, the calculation 
using the number of potential moves (potential move) and the direct calculation 
with $H(n_b)$ (direct).}

\label{table1}
\begin{ruledtabular}
\begin{tabular}{lll}
$\overline{\epsilon (T)}$    & potential move & direct \\
\hline
cluster-flip flat histogram  & 0.0002  & $\cdots \cdots$ \\
multibondic                  & 0.0002  & 0.043      \\
improve multibondic          & 0.0002  & 0.0087     \\
\end{tabular}
\end{ruledtabular}
\end{table}

\section{Summary and discussions}

To summarize, we have derived the rigorous BHR for the bond number, 
investigating the cluster (graph) representation of the spin models.  
We have shown that the bond-number DOS $\Omega(n_b)$ can be calculated 
in terms of $\left< n_p \right>_{n_b}$.   We have proposed 
a Monte Carlo dynamics based on the number of potential moves 
for the bond number, which is regarded 
as the flat histogram method for the bond number.  
We have shown the efficiency of the BHR for the bond number 
in calculating the bond-number DOS and other physical quantities. 

For the dynamics, the combination of the Wang-Landau idea \cite{WL01} 
and the cluster algorithms is useful in accelerating the diffusion
of the random walker, as was pointed out before \cite{Yama02}.  
However, here we have made more emphasis on the use of the 
BHR for the estimator of $\Omega(n_b)$. 
The advantage of using the BHR may be attributed to the fact that 
the number of potential moves is a macroscopic quantity, which is the same 
situation as the BHR for the energy \cite{Oliv00,Lima00}.  
Moreover, the use of the improved estimator for calculating 
the number of potential moves, Eq.~(\ref{eq:Ocal2}), 
gives much better statistics for the calculation. 

The number of potential moves for the energy, $N(S, E \to E \pm \Delta E)$, 
has several possibilities for $\Delta E$.  
On the contrary, in the case of the number of potential moves 
for the bond number, $N(G, n_b \to n_b \pm 1)$, the change of 
the bond number is limited to one, which makes 
the calculation of the bond-number DOS through the number of potential moves 
much simpler than that of the energy DOS. 

Recently, a cluster Monte Carlo algorithm to simulate the $Q$-state 
Potts model for any real $Q \, (>0)$ was proposed by Gliozzi \cite{Gliozzi}.  
It is interesting to apply the BHR to that method.   
Since only the information on graph is used in that Monte Carlo algorithm, 
Eq.~(\ref{eq:np2}) is useful for calculating $\left< n_p \right>_{n_b}$. 

In this paper, we argued the BHR for the bond number. 
We can extend the present idea to the relation including two variables, 
for example, the bond number and the cluster number.
The extension to more general cases, such as the loop algorithm of 
the quantum Monte Carlo simulation, may attract much attention, 
which will be studied in near future. 

\section*{Acknowledgments}

We thank H. Otsuka, Y. Tomita, and J.-S. Wang for valuable discussions.
This work was supported by a Grant-in-Aid for Scientific Research 
from the Japan Society for the Promotion of Science.


\end{document}